\documentclass[letterpaper]{IEEEtran}
\pdfoutput=1
\usepackage{comment}
\usepackage{graphicx}
\usepackage{multirow}
\usepackage{epstopdf}
\usepackage{bm}
\usepackage{tabularx}
\usepackage{mathrsfs}
\usepackage{url}
\usepackage[nospace,noadjust]{cite}
\usepackage{array}
\usepackage{diagbox}
\usepackage{amssymb}
\usepackage{mdwmath,amsmath}
\usepackage{color}
\usepackage{balance}
\usepackage[caption=false,font=footnotesize]{subfig}
\usepackage{balance}  % to better equalize the last page
\usepackage{graphics} % for EPS, load graphicx instead
\usepackage{times}    % comment if you want LaTeX's default font
\usepackage{mathptmx}
\usepackage{leading}
\usepackage{pifont}
\usepackage{amsthm}
\usepackage[normalem]{ulem}
\usepackage[top=1.5in, bottom=1.5in, left=1in, right=1in]{geometry}
\usepackage{pbox}

\usepackage[caption=false,font=footnotesize]{subfig}

\DeclareMathOperator*{\argmax}{argmax}

\graphicspath{{fig/}}

\DeclareRobustCommand*{\IEEEauthorrefmark}[1]{%
  \raisebox{0pt}[0pt][0pt]{\textsuperscript{\footnotesize\ensuremath{#1}}}}

\begin{document}

%\title{Towards Characterizing Truck Turning Behaviors with Application to Safety Driving}

\title{Vehicle Traffic Driven Camera Placement for Better Metropolis Security Surveillance }
\author{\IEEEauthorblockN{
Yihui He\IEEEauthorrefmark{1}$^*$,
Xiaobo Ma\IEEEauthorrefmark{1}\IEEEauthorrefmark{2}$^*$\thanks{$^*$Equal contribution.},
Xiapu Luo\IEEEauthorrefmark{1}\IEEEauthorrefmark{2}$^\dagger$\thanks{$^\dagger$Corresponding author.},
Jianfeng Li\IEEEauthorrefmark{1}\IEEEauthorrefmark{2},
Mengchen Zhao\IEEEauthorrefmark{3},
Bo An\IEEEauthorrefmark{3},
Xiaohong Guan\IEEEauthorrefmark{1}\\
%,~\IEEEmembership{Fellow, IEEE}\\
\IEEEauthorblockA{\IEEEauthorrefmark{1}MOEKLINNS Lab, Xi'an Jiaotong University, Xi'an, China}\\
\IEEEauthorblockA{\IEEEauthorrefmark{2}Department of Computing, The Hong Kong Polytechnic University, Hong Kong}\\
\IEEEauthorblockA{\IEEEauthorrefmark{3}School of Computer Science and Engineering, Nanyang Technological University, Singapore}}

%{Email: 
%\IEEEauthorrefmark{2}\{jfli, jtao, xbma\}@sei.xjtu.edu.cn, junjie.zhang@wright.edu, xhguan@sei.xjtu.edu.cn}

}

\maketitle
\thispagestyle{plain}
\pagestyle{plain}

\begin{abstract}

Security surveillance is one of 
the most important issues    in  smart cities, 
especially  in an era of   terrorism.
Deploying a number of (video) cameras is a common surveillance approach.
 Given 
 the never-ending   power offered by vehicles to   metropolises, 
exploiting vehicle traffic to design  camera placement strategies
could potentially facilitate     security surveillance.
This article   constitutes the first effort toward building the linkage between vehicle traffic and security surveillance, which  is a critical problem for  smart cities. 
We expect our study 
could  
influence the decision making  of surveillance camera placement, 
and  foster   more research of principled ways of security surveillance beneficial to our physical-world life. Code has been made publicly available\footnote{github.com/yihui-he/Vehicle-Traffic-Driven-Camera-Placement}.

\end{abstract}

%\begin{IEEEkeywords}
%  camera placement, security surveillance, vehicle traffic
%\end{IEEEkeywords}

\section{\textbf{Introduction}}
\label{sec:intro}
Security surveillance is one of
the most important issues   in   smart cities.
Due to  the continuous  growth of cities   in   size and complexity, 
keeping cities safe becomes   critical to attracting    skilled people and investments  necessary for economic growth and development. 
Compounded by  terrorism,
  cities, especially  metropolises,  have to carefully conduct security surveillance.

To fight against the adversary,
 deploying a number of (video) cameras is a common surveillance approach, which  has gained prominence in policy proposals on combating
terrorism~\cite{camera_terrorism}.
At first glance, camera placement is a trivial task
that could be achieved by simply purchasing a huge number of cameras.
However,
subject to the economic budget, 
cameras  cannot  be placed unlimitedly. 
Additionally, as the camera resolution increases,
the  network bandwidth (and the storage and processing overhead)  needed for real-time transmission of the recorded videos to the data center is considerably large. 
For example, the latest generation of IP cameras 
require up to  10  Mbps bandwidth per camera~\cite{band}.
%(frame rate 5 FPS, stream type MJEPG,   resolution 1080P HD, and high video quality). 
Theses constraints will  become more significant as cities grow in
size and complexity.

%Given 
% the never-ending  power offered by vehicles  to   metropolises, 
Exploiting vehicle traffic  from cyber transportation  space to strategically place cameras
could potentially facilitate  physical-world  security surveillance. 
For example, criminals and terrorists  may  drive vehicles before and after security events occur; vehicles have become   terrorists' favorable  weapons, such as
the 2017 Westminster attack~\cite{attackc}. Thus,
monitoring vehicles has  great potential  in collecting security event evidence. 
%Fortunately,   the prosperity of Internet of Things  
%makes it mature to monitor  vehicle traffic via  GPS sensors, 
% opening up   new opportunities
%  to reconsider   surveillance   by performing vehicle traffic driven camera placement. 
%   Careful planning of cameras would result in effective surveillance,
%      enabling more responsive dispatch of  security forces where
%and when   needed, and consequently improving security. 

As a straightforward strategy, one may place cameras in     busy (e.g., more traffic) regions, where the adversary 
tends to commit crimes  or   large-scale  threatening terrorism activities.
However, this strategy may fail to
cover   vehicles that never enter busy regions. 
Intuitively, a good   strategy should allow 
 distinct vehicles visible to cameras as many as possible,
 and make each vehicle  exposed to  cameras frequently.
 The former requirement
is very important because
  the  invisibility of a vehicle to cameras means
  the lack of surveillance. 
  The latter requirement encourages more information to be collected for each vehicle, 
  making it more likely to  recognize the trajectory
  of a vehicle.
 % These two requirements, however, may conflict each other. 

After  placing the   camera infrastructure, 
we also expect the placed cameras to be of high resolution for high-quality video data.    
However, 
  as the    number of placed cameras  grows, 
 the overhead  of handling the data (e.g., transmission, storage and processing) may become intractable.
That is, despite the large population of placed cameras, 
   the number of   cameras that could be of high resolution simultaneously is limited. 
Determining which cameras should be of high resolution is a  non-trivial task. If a fixed set of cameras are configured as high resolution, the adversary may evade these cameras (when committing crimes), thereby eliminating high-quality crime evidence.

Focusing on these problems, this article   investigates
 vehicle traffic driven security camera infrastructure placement, and  formulates  it as a group of submodular set function optimization problems, which could be gracefully  solved by the greedy algorithm with  theoretical lower bounds.
We propose   five different placement strategies that differ in designing goals. 
Also, we    design a randomized camera resolution upgrading framework based on game theory,
with the purpose of breaching the practical  overhead barrier of  handing video data.   Using the carefully designed   randomized strategy,  we can  
 deter  the sophisticated  adversary with maximal utility.

%We finally perform experiments to analyze  security surveillance  
%from different aspects  using real-world   vehicle traffic.

%The contributions are summarized below. 
%\begin{itemize}
%\item
 Our work constitutes the first effort toward building the linkage between vehicle traffic and security surveillance, which  is a critical problem for  smart cities. 
% We proposed five strategies that are all formulated to meet the submodular property, thereby easy to be solved with   theoretical lower bounds. 
%\item To   our best knowledge, our proposed strategies, especially $S3 \sim S5$, are novel. 
%The novelty lies in that $S3 \sim S5$ not only encourage the number of vehicles under surveillance, but also stimulate more (unique) camera hits and in-camera time for each vehicle. 
%\item We compare and analyze the proposed strategies based on different metrics, such as unique vehicle coverage ratio and vehicle traffic coverage ratio, using real-world vehicle traces in a large metropolis. Also, we evaluate the possibility of recovering the invisible trajectory of a vehicle between two successive cameras under different strategies. 
%\item We consider the security surveillance scenario where one (e.g., security police) wants to collect  qualified video crime evidence, while only a limited number of placed cameras can be of high clarity due to   overhead barrier.   We formulate this problem as a Stackelberg security game, and solve it with the linear programming technique. Experiments demonstrate that the derived mixed strategy (i.e., randomized strategy following a probability distribution)   can maximize the surveillance utility against a sophisticated (i.e., best-response) adversary.  
%\end{itemize}
%
%
%
We expect our study 
could  
influence the decision making  of surveillance camera placement, 
and  foster   more research efforts towards principled ways of security surveillance beneficial to our physical-world life.

\section{\textbf{Vehicle Traffic Driven Placement Strategies for Building Surveillance Infrastructure}}

\label{sec:detect}
\subsection{\textbf{Problem Description}}
Let $V$ be the set of vehicles   under surveillance and  $v \in V$ be a vehicle. 
Suppose   $v$ is equipped with a position sensor.
%The sensor records the sequence of its position coordinates  with a  time interval $t$. 
%which could 
%be constant or variable. 
The coordinates include latitude and  longitude.
%,   denoted by $x$ and $y$,  respectively.
Specifically,  the data from the  sensor  
comprises the following fields: (\textit{vehicle ID}, \textit{time}, \textit{latitude}, \textit{longitude}), 
where the vehicle ID    uniquely identifies a vehicle.

%Prior to  placing  cameras, 
The   metropolis is characterized as  a rectangle  uniquely identified by   
its maximum latitude, minimum latitude, maximum longitude, and minimum longitude.
%  the maximum latitude ($x_\text{max}$), minimum latitude ($x_\text{min}$), maximum longitude ($y_\text{max}$) and minimum longitude ($y_\text{min}$) 
%  of $M$. 
  The rectangle is  
divided   into small blocks with a size of
$l \times l$ (e.g., $50m\times50m$), resulting in
  a set of blocks   denoted by $\Omega$.
 The advantages of dividing the map into blocks are two-fold, namely, map-independent (applicable for all metropolises) and exactly   covering the entire map (unlike circulars).
%Note the   regions in $\Omega$  at the edge of metropolis $M$ may not be
%with  a size of
%$l \times l$ since the length or width the rectangle  may not be divisible by $ l$.
 
Our goal is 
to find a subset of   blocks $C \subseteq \Omega$ to place (super) cameras, so that the security surveillance utility could be maximized. The  cardinality of $C$ is the number of blocks  that we select to place cameras. Let $N$ denote the maximum  number of blocks where we can afford to place cameras. Obviously,
we have $|C| \leq N$. 
Note that only a block  with GPS records can be selected for camera placement, thereby containing at least one road.
%A good placement strategy should
%select blocks that are  informative. Specifically, deploying cameras  should benefit  monitoring more unique vehicles and meanwhile observing each vehicle frequently. The former requirement
% could raise the possibility of discovering at least  one block where 
% a vehicle appears.
% This is of crucial importance because
% the  invisibility of a vehicle to cameras means
% the lack of surveillance. 
% The latter requirement encourages more information to be collected for each vehicle,  
% making it more likely to recognize the trajectory
% of each vehicle.
% %, which further enriches the surveillance information.
%Obviously, different placement strategies differ in objective functions, yielding  different subsets of  blocks $C$. 
%%to place cameras.

\subsection{\textbf{Placement Strategies}}
%We   propose five   vehicle traffic driven camera placement strategies 
%that facilitate security surveillance from different aspects. 
For all strategies, 
we denote the  objective   function  by $F(C)$.

\subsubsection{\underline{S1}---Maximum Unique Vehicles}

This strategy selects a subset $C \subseteq \Omega$ so 
that the number of unique vehicles crossing (at least one)     blocks
in $C$ is maximized, formally   expressed as
\begin{equation}
\label{eq1}
\argmax_C F(C) = \sum_{v\in V} I_v(C)
\end{equation}
where $I_v(C)$ equals one if  $v$ 
has  crossed at least one  block in $C$; otherwise $I_v(C)$ equals 0.
Such a strategy  maximizes the visibility space (i.e., the total number of unique vehicles) of security surveillance.

\subsubsection{\underline{S2}---Maximum  Vehicle Traffic}

This strategy differs from S1 in that 
it maximizes the amount of vehicle traffic (rather than unique vehicles) crossing    blocks
in $C$. Similarly, it is
\begin{equation}
\label{eq2}
\argmax_C  F(C) =\sum_{v\in V} T_v(C)
\end{equation}
where $T_v(C)$ denotes 
the amount of traffic    of blocks in $C$ contributed by
 $v$, and could be measured by  the total time when $v$ stays inside $C$.
%This signifies that a vehicle staying for
%a longer time  in $C$
%contributes more to the 
%traffic in $C$.
 $T_v(C)$ can  be calculated by  $\sum_{c \in C} T_v(c)$,
where $T_v(c)$ denotes the total time when $v$ stays in $c$, depending on the times when $v$ enters and leaves $c$. 
%
%In particular,
%when the position records are generate at a regular time interval,
%$T_v(C)$
% could be calculated 
%as the total number of position records generated when $v$ is inside $C$.
%
S2 maximizes the total visible time to cameras of all vehicles (i.e., summing up the time  each vehicle under surveillance).

\subsubsection{\underline{S3}---Minimum Mean OITR  (Out-Camera  to In-Camera Time Ratio)}
%Another strategy  in favor of security surveillance
%is to 
This strategy minimizes the mean 
OITR across all vehicles. 
The OITR of a vehicle 
represents the proportion of  time  when a vehicle is outside the visible scope of cameras. 
%ACIs could also be approximately considered as the average duration 
%when a vehicle is invisible to cameras. 
Intuitively, 
smaller mean OITR across all vehicles indicates better security surveillance.
S3   is equivalent  to the  maximization problem:
\begin{equation}
\label{eq3}
\argmax_C F(C)=\Phi-\sum_{v \in V}  \left(\frac{S}{T_v(C)+1}-1\right)/{|V|}
\end{equation}
where $S$ is the   measurement time. 
  S3 does not take   the uniqueness of
cameras into consideration. Therefore, it encourages more unique vehicles    visible to (whichever) cameras, and meanwhile each   vehicle visible to cameras as long as possible.
\textit{Note that $T_v(C)$ 
is increased by one  to avoid $S$ being divided by zero, and $\Phi$ is a constant derived from  $F(\emptyset)=0$ so that we have $F(C)\geq 0$. The same operation exists for  S4  and  S5.}

\subsubsection{\underline{S4}---Minimum Mean ACIs (Average Camera-Hit Intervals)}
We define the ACIs   to 
represent  the average time interval to hit a 　 camera  for a vehicle.  Formally, we minimize the 
mean ACIs across all vehicles, and 
$S4$ is equivalent to:
\begin{equation}
\label{eq4}
\argmax_C F(C)=\Phi- \sum_{v \in V}  \frac{S}{H_v(C)+1}/{|V|}
\end{equation}
where $H_v(C)$ denotes the number of times  
  that vehicle $v$ hits (whichever) cameras  along its trajectory during the 
    measurement time  $S$.
% $H_v(C)$ is calculated as the number of times that 
%regions in $C$ that  $v$ goes across. 
%Similarly, $H_v(C)$ 
% is increased by one  to avoid  zero denominators, and $\Phi$ is a constant derived from  $F(\emptyset)=0$ so that we have $F(C)\geq 0$.
S4   encourages more unique vehicles
visible to (whichever) cameras, and meanwhile each   vehicle to hit cameras more frequently.

\subsubsection{\underline{S5}---Minimum Mean AUIs (Average Unique-Camera-Hit Intervals)}
We further adapt   $S4$ by 
  considering the uniqueness of cameras that a vehicle  hits. 
Accordingly, we define the AUIs   to 
represent  the average time interval to hit a new camera for a vehicle. Similar to S4, we  minimize the 
mean AUIs across all vehicles, and 
S5     is defined as: 
\begin{equation}
\label{eq5}
\argmax_C F(C)=\Phi-  \sum_{v \in V}  \frac{S}{U_v(C)+1}/{|V|}
\end{equation}
where $U_v(C)$ denotes the number of 
unique cameras that vehicle $v$ hits during the measurement time $S$.
$U_v(C)$ is calculated as the number of unique  
blocks in $C$ that   $v$ crosses. 
%Again, $U_v(C)$ 
% is increased by one  to avoid  zero denominators, and $\Phi$ is a constant derived from  $F(\emptyset)=0$ so that we have $F(C)\geq 0$.
S5   encourages
more unique vehicles
visible to (whichever) cameras, and meanwhile each   vehicle to hit more new cameras.

\subsection{\textbf{Solving Optimal Placement Strategies}}

All the above   strategies are formulated as   maximization problems under the constraint   $|C| \leq N$. To find a subset of blocks $C \subseteq \Omega$ to place cameras so that $F(C)$ can be maximized,
we need to solve these maximization problems, which are  NP-hard.

Each strategy differs in the  objective  function $F(C)$.
However, this does not necessarily mean that
we have to solve these  strategies in five different manners. Instead,  all these 
maximization problems could be gracefully solved 
with a greedy algorithm (GA) 
because of their non-increasing monotony and  submodularity~\cite{citeulike:416655, krause2012submodular}.
The non-increasing monotony 
means   that, for any two sets $C_1, C_2 \subseteq \Omega$ and $C_1 \subseteq C_2$,  we have $F(C_1) \leq F(C_2)$.
Apparently, the functions defined in {S1}$\sim${S5} are non-decreasing.
%, which is not hard to prove for the functions defined in equations  from (1) to (5).
%, which is not hard  to prove for our strategies and thus we omit the proof.

The submodularity    means that a  non-decreasing set function has the property of diminishing returns when a single element $c$ is added to an input set $C$, as compared to $c$ is added to an input set that is a subset of $C$. 
%A
%submodular function has the    property that the difference in the incremental value of the function, that a single element makes when added to an input set, decreases as the   input  increases. 
Specifically, for a non-decreasing function, such as $F(C)$ defined in  {S1}$\sim${S5},
given every two sets $C_1, C_2 \subseteq \Omega$ and $C_1 \subseteq C_2$, and a new block $c \in \Omega \setminus C_2$,
the submodular property is equivalent to
$F(C_1 \cup \{c\})-F(C_1) \geq F(C_2 \cup \{c\})-F(C_2)$.
This means smaller sets have more function value increment when they are added with a new block.
It is easy to prove the the submodularity for S1 and S2.  The proof of the submodularity for {S3}$\sim${S5}
is included in the appendix.
Interested readers are referred to~\cite{krause2012submodular} for  a comprehensive survey of submodularity.
% \cite{extension}.

According to~\cite{citeulike:416655}, submodularity allows us to 
%to obtain an
%approximate solution that is at least $1- 1/e \approx  63 \%$ of the
%optimal solution.
%In other words, GA is able to,  for  all these
%problems,  
derive a solution   lower bounded 
by $1- 1/e \approx  63 \%$ of the optimal solution.
GA runs for at most $N$ rounds
to obtain a set $C$ of size $|C| \leq N$.
In each round, it selects a new  block
$c \in \Omega \setminus C$   maximizing the reward gain, $\delta_c(C)=F(C\cup{c})-F(C)$,
and inserts $c$ into $C$.
This process repeats   until $|C|=N$ or  $\delta_c(C)=0$. 

%To be more precise,
%the reason that GA is quite fit for our problems 
%is the common property that $F(C)$ possesses
%across all strategies: non-decreasing 
%submodular  functions.

%
%%We omit the proof the summodular property of $F(C)$ in the proposed strategies. 
%Note that 
%$F(C)$ in   \eqref{eq2} 
%could reach a global optimal solution using  GA.
%% under the constraint of $|C| \leq N$.
%Although  
%$F(C)$ in   \eqref{eq2}  complies with the submodularity, 
%%its submodular property 
%% can be  more  concrete. That is,  
% given every two sets $C_1, C_2 \subseteq \Omega$ and $C_1 \subseteq C_2$, and a new block $c \in \Omega \setminus C_2$,
% the submodularity concretely means
% $F(C_1 \cup \{c\})-F(C_1) = F(C_2 \cup \{c\})-F(C_2)$.
% This indicates that 
%adding   a new  block
%yields equal function value increment. 
%In this case,  using GA to solve $F(C)$ in  \eqref{eq2}
%results in a global  optimal solution, i.e., 
%  top $N$   blocks regarding vehicle traffic. 
%
%%
%We finally present the greedy solving procedure.

%The computational complexity of this algorithm is $O(N)$

\section{\textbf{Randomizing Camera Resolution Upgrading for High-quality Surveillance}}
\label{sec_model}

As the   number of the \textit{affordable} cameras  grows, 
 the overhead  of handling the videos (e.g., transmission, storage and processing) may be impossible to satisfy,
especially when  high-resolution video is desired for  high-quality crime evidence~\cite{Fularz2015}. 
In other words,  only a limited number of   cameras could be of high resolution  simultaneously. 

Determining which cameras should be of high resolution is  non-trivial. 
If a fixed set of cameras are configured as high resolution, the adversary may 
evade  them deliberately (when committing crimes), thereby 
eliminating high-quality crime evidence.
Therefore, we propose to randomly choose a set of placed cameras and upgrade their resolution, with an attempt to deter the adversary by making him feel that any camera  might be of high resolution. 
%Such randomization requires the remote configuration of camera clarity, which 
%is technically feasible given the vast availability of smart IP cameras~\cite{ipcamera, 7095560}.   

%The basic idea is to randomly activate and deactivate cameras, and apply the randomized policy in the camera infrastructure to deter the adversary by making him feel that all the deployed cameras might be working. The reasonability of this idea is two-fold. First, as metropolises develop, there is a growing tendency that the affordable number of cameras becomes extremely large. Second, the latest cameras may require up to 10Mbps bandwidth due to their higher resolution and greater clarity, which is very beneficial to collecting more qualified video-based crime evidence. As a consequence, the transmission, the storage and the computation overhead of dealing with the video would be extremely high.  
%Therefore, we propose a randomized camera on-off strategy  to deter criminals with limited overhead.

\subsection{A Game-theoretic Formulation}
Since   different blocks    differ in their priorities to be monitored, or to be the target blocks where criminal activities are committed, we take into account the importance   of different blocks when
designing the randomized strategy. 
Meanwhile,  
the adversary aims to evade the  surveillance of high-resolution cameras, while the defender (e.g.,  police)  tries to 
observe the adversary. 

To describe such confrontation, we 
leverage
  the Stackelberg security game  (SSG) to 
   design the strategy~\cite{Clempner:2015:SSG:2774253.2774361}.   SSG is well-suited to adversarial reasoning for security resource allocation  problems and has been adopted in many applications, such as  US Federal Air Marshal Service~\cite{ssg}. 
   %It can be naturally used to model real-life security surveillance problems.

A standard SSG has two players, a
leader and a
follower. Each player has its own   set of 
pure strategies to select. The leader and the follower act sequentially as follows.

\noindent
\textbf{Step 1.}  The leader (i.e., defender) commits to
a mixed strategy.
The
mixed strategy
allows the leader to play a probability distribution over
pure strategies, maximizing 
the leader's utility. 

%The leader's  strategy is designed in consideration of the follower's response model, 

\noindent
\textbf{Step 2.} The follower (i.e., adversary), as a response,  selects a pure strategy that optimizes his utility after inspecting and  learning  the mixed strategy chosen by the leader.

%The follower's strategy takes into account the learned mixed strategy of the leader.      

We consider  a threat model wherein the adversary is \textit{sophisticated}. Specifically,  
the adversary     can  inspect and   learn the monitor's mixed strategy (i.e., the probability distribution), and 
select the pure strategy that maximizes his utility, i.e., a best response adversary.

%strong and rational. This means that i) the adversary     can \emph{perfectly} inspect and   learn the monitor's mixed strategy (i.e., the probability distribution);　 ii) and 
%the adversary selects the pure strategy that maximizes his utility (i.e., best-response adversary). 
On the other hand,
the defender  is forward-looking. That is, she
takes into account     the   adversary's threat model when designing strategies, thereby making her strategy robust against the sophisticated adversary.

Although the adversary can learn the defender's mixed strategy, 
he cannot predict  which specific pure strategy the defender would adopt at the time of his scheduled criminal activities.

\subsection{Player   Strategies}

A pure strategy of the defender is  a set of cameras whose resolution can be upgraded  simultaneously. 
  Our aim is to deter the adversary by randomly upgrading the resolution of
 the placed cameras, and thus the adversary committing crimes in blocks  without cameras 
 are not considered. 
 Therefore, a pure strategy of the adversary is a set of blocks  with  placed cameras.

\subsection{Utility Functions}

Consider that the adversary commits crimes  in the $i$th block $c_i$. 
If $c_i$ is covered by the defender's pure strategy, the  defender 
receives
reward $R_i^d$. Otherwise, the defender receives penalty $P_i^d$.
Similarly, the adversary receives penalty
$P_i^a$ in the former case, and reward $R_i^a$ in the latter case.

The reward   that the defender monitors  a specific block  can be
assigned according to the importance  of  the block, such as the amount of vehicle traffic, the number of unique vehicles, and 
 historical crime activity severity,   and so forth.  
The  penalty that     the adversary selects   a block can be measured in the same way.

Let   $\Gamma_j$ denote  the $j$th  defender pure   strategy,  and
$A_{ij}$ denotes the coverage indicator of $\Gamma_j$ on $c_i$, where
 $A_{ij}=1$ for $c_i \in \Gamma_j$, and  $A_{ij}=0$ for $c_i \notin 
 \Gamma_j$. Let $J$ be the number of defender pure strategies. 
The number of adversary pure    strategies  is $N$, the (maximum) number of blocks where cameras are placed.
%, since we consider  a pure strategy of the adversary as each block where to commit crimes.    
 We denote the 
 probability of the defender choosing pure  strategy $\Gamma_j$   by $a_j$, 
 and we have 
\begin{equation}
\sum_{j=1}^J a_j=1
\end{equation} 
  The marginal  probability $x_i$ for the defender to upgrade  $c_i$ (i.e., upgrade the resolution of the camera in   $c_i$)
can be calculated by
\begin{equation}
\setlength{\abovedisplayskip}{3pt}
\setlength{\belowdisplayskip}{3pt}
%\small
	x_i=\sum_{j=1}^J a_j A_{ij}, i=1,2\ldots,N
\end{equation}
%The summation of $x_i$ denotes the expected number of
%blocks that the defender can upgrade simultaneously, and has an upper bound of $|\Gamma|_\text{max}$, i.e.,  the maximum number of cameras covered by $\Gamma_j \text{,} \forall j$.
We denote $(a_1,a_2,\ldots,a_J)$ by $\mathbf{a}$, and
 $(x_1,x_2,\ldots,x_{N})$ by $\mathbf{x}$, where $\mathbf{x}$ is determined by $\mathbf{a}$.

The defender's expected utility on  upgrading $c_i$
can be expressed as
\begin{equation}
\label{uid}
\setlength{\abovedisplayskip}{3pt}
\setlength{\belowdisplayskip}{3pt}
%\small
\begin{split}
	U_i^d(x_i) &= x_i R_i^d   + (1-x_i)P_i^d
		\end{split}
\end{equation}
and   the adversary's expected utility on committing crimes in $c_i$ is:
\begin{equation}
\label{uia}
\setlength{\abovedisplayskip}{3pt}
\setlength{\belowdisplayskip}{3pt}
%\small
\begin{split}
	U_i^a(x_i) &= x_i P_i^a  + (1-x_i)R_i^a
	\end{split}
\end{equation}

\subsection{Mixed Strategy against A Best Response Adversary}
\label{sec_br}
We formally define the objective function
with constraints to maximize the defender's utility  against
a best response adversary. 
%With the pure strategies and utility 
%functions, the defender will design her \emph{mixed} strategy 
%(i.e., $\mathbf{a}$) maximizing the defender's utility. 
%Since the defender's utility takes into consideration of the adversary's 
%responses (i.e., how the adversary with the knowledge of the 
%defender's mixed strategy performs attacks), 
%solving the optimal mixed   strategy relies on the adversary's 
%response models.
Let us first consider a sophisticated adversary who   takes the best response, i.e., 
  selecting the block to commit crimes so  to maximize his   utility.  In this case,
the probability that the adversary
selects $c_i$
equals
\begin{equation}
\label{br}
%\small
\setlength{\abovedisplayskip}{3pt}
\setlength{\belowdisplayskip}{3pt}
B_i   =
\begin{cases}
1 & U_i^a(x_i) \geq U_j^a(x_j), \forall j=1,\ldots,J\\
0 & \text{otherwise}
\end{cases}
\end{equation}
This means that the adversary
\textit{knows} the marginal probability $x_i$ for the defender to upgrade   $c_i$, and he
 selects the target with maximal expected utility.
 Consequently, the utility  functions of the adversary and the defender can be expressed as
 \begin{equation}
\label{object_aduti}
\setlength{\abovedisplayskip}{3pt}
\setlength{\belowdisplayskip}{3pt}
%\small
  U^a= \sum_{i=1}^{N} B_i U_i^a(x_i)
\end{equation}
  \begin{equation}
\label{object_deuti}
\setlength{\abovedisplayskip}{3pt}
\setlength{\belowdisplayskip}{3pt}
%\small
  U^d= \sum_{i=1}^{N} B_i U_i^d(x_i)
\end{equation}
%In case there are multiple targets giving the adversary the same highest utility, the adversary will attack the target that  gives the defender the highest utility, as the defender can induce the attacker to do so by changing the strategy with a very small amount.

Simultaneously, the defender selects an optimal mixed (i.e., randomized) strategy in consideration of   the sophisticated adversary's best response.
%, where the adversary breaks ties in  favor of the defender\cite{Korzhyk:2011:SVN:2051237.2051246}.
The defender maximizes her   utility $U^d$   as
\begin{equation}
\label{object_br}
\setlength{\abovedisplayskip}{3pt}
\setlength{\belowdisplayskip}{3pt}
%\small
\max_{\mathbf{a}}   \sum_{i=1}^{N} B_i U_i^d(x_i)
\end{equation}
Substituting    \eqref{uid}, we rewrite   \eqref{object_br} as
\begin{equation}
\label{object_br_rw}
\setlength{\abovedisplayskip}{3pt}
\setlength{\belowdisplayskip}{3pt}
%\small
\max_{\mathbf{a}}  \sum_{i=1}^{N} B_i \left(x_i R_i^d   + (1-x_i)P_i^d\right)
\end{equation}

%In the LFA detection problem,
%there are constraints for the defender to assign   security resources.
%As interpreted in previous sections,
%not all possible subsets of links can be feasible
%strategies of the defender,
%and the marginal coverage ratio $x_i$ on target $l_i$ is also restricted.
%Actually,
 %$x_i$ can be calculated
%by $x_i=\sum_{j=1}^J a_j A_{ij}$,
%where $a_j$ is the probability of
%choosing the defense strategy $\Gamma_j$ and $A_{ij}$ is coverage indicator of $\Gamma_j$ on $t_i$ (i.e.,  $A_{ij}=1$ if $t_i$ is covered by $\Gamma_j$; $A_{ij}=0$ otherwise).
To calculate the defender's optimal   strategy, the following  problem $\text{P1}$ needs to be solved.
\begin{equation}
\label{object_function}
\setlength{\abovedisplayskip}{3pt}
\setlength{\belowdisplayskip}{3pt}
%\small
\textbf{P1:}\!\left\{
\begin{split}
&	\max_{\mathbf{a}}  \sum_{i=1}^{N} B_i \left(x_i R_i^d   + (1-x_i)P_i^d\right)  \\
& \text{subject to   } \hspace{15mm}  x_i=\sum_{j=1}^J a_j A_{ij}\text{, } \forall i\\
%& \hspace{30mm} \sum_{i=1}^I x_i \leq |\Gamma|_\text{max}\\
& \hspace{30mm} \sum_{j=1}^J a_j = 1\\
& \hspace{30mm} 0 \leq a_j \leq 1 \text{, } \forall j
\end{split}	
\right.
\end{equation}
Problem P1    cannot be directly solved since $B_i$ defined in \eqref{br}
contains an underlying ``if-then'' logical relationship. Instead, after removing the ``if-then'' logical relationship,  P1 can be   readily solved by the branch-and-cut algorithm~\cite{Gavalas2014}.

The  defender finally
   adopts the mixed strategy below.
   
\begin{equation*}
\small
\setlength{\abovedisplayskip}{3pt}
\setlength{\belowdisplayskip}{3pt}
\boxed{
\begin{aligned}
&\text{Mixed Strategy:   }  \text{play } \Gamma_j \text{with probability } a_j \hspace{10mm}\\
&\hspace{7mm} \text{where  } a_j\in \mathbf{a} \text{ is the solution 
of problem P1}\\ 
%& \hspace{10mm}\text{utility (drived from problem P1).} \\ 
& \hspace{7mm}j=1,2,\ldots,J
%   \\
%& \hspace{7mm} 
%j=1,2,\ldots,J
\end{aligned}
}
\end{equation*}

\section{\textbf{Experimental Evaluation}}
%We evaluate the proposed strategies using  taxi data from    Beijing, a metropolis with a population of more than 20 million permanent residents and an area of more than 15,000 square kilometers, to place cameras for  performing taxi-specific   security surveillance.

\subsection{\textbf{Dataset Description}}
The   data contains one-week  GPS  trajectories  of  10,357  taxis
in Beijing, with more than  15 million position records and  9 million kilometers total distance of the trajectories~\cite{data}. 
 We remove the outliers  along the GPS trajectories of a vehicle, including points indicating an impossible speed,
 and points that significantly deviate the moving average. 
%First, for any given GPS point, we calculate the speed by comparing the displacement and the time interval to the GPS point anterior to it. If the speed exceeds the speed threshold, the GPS point will be removed. This procedure repeats along the sequence of the GPS points of a vehicle until all the GPS points (except the first one) are examined. We specify the threshold equal to 200 kilometers per hour, as it is unlikely for a taxi in Beijing to run so fast. 
%Second, given a GPS point  $ ({x}_i, {y}_i)$, we calculate the average speed along the $x$ dimension and the $y$ dimension respectively, by taking the last $n$ points before  $ ({x}_i, {y}_i)$. Then, we apply the calculated average speed to the last point before  $ ({x}_i, {y}_i)$ so to get the predicted location of  $ ({x}_i, {y}_i)$, denoted as  $ (\hat{x}_i, \hat{y}_i)$. If the distance between  $ ({x}_i, {y}_i)$ and $ (\hat{x}_i, \hat{y}_i)$  does not exceed the distance threshold, then  $({x}_i, {y}_i)$ would be blended with   $ (\hat{x}_i, \hat{y}_i)$  by  $ ({x}_i, {y}_i) = \alpha ({x}_i, {y}_i) + (1-\alpha) (\hat{x}_i, \hat{y}_i)$, where  $ 0<\alpha<1$. Otherwise, we consider    $({x}_i, {y}_i)$ as an outlier, remove it and substitute it by  $ (\hat{x}_i, \hat{y}_i)$. We specify $n$  as 10, the distance threshold as 50 meters, and  $\alpha$ as 0.9 to put more weight to the observed GPS points. We continue apply this procedure to the GPS points following $({x}_i, {y}_i)$. 
Finally, we  identify  0.28\% of all the GPS points as outliers.
%  The average sampling interval is about 177 seconds with a distancence
%of about 623 meters. 

% Beijing ranges from $39.4^\circ$   to $41.6^\circ$ N in latitude, and  from $115.7^\circ$  to $117.4^\circ$ E in longitude. Accordingly, 
We then divide Beijing into a set of   blocks $\Omega$   with a size of
  $50m\times50m$, yielding the total number of such blocks $|\Omega|=14,473,089$.  
 We can place cameras in  blocks 
 where vehicles arrive.
 % (i.e., the blocks where GPS records exist).
  There are 438,674 such  blocks, denoted as $R\subset \Omega$.

\subsection{\textbf{Performance   of Camera Placement  Strategies}}
\subsubsection{Metrics}
Larger values of these metrics indicate better security surveillance.
% from different aspects.
\begin{itemize}
\item {\underline{UCR}  (Unique Vehicle Coverage Ratio)}. The ratio of  observed   unique vehicles 
  to all vehicles,  calculated by
$\sum_{v \in V}I_v(C)/\sum_{v \in V}I_v(R)$.
\item {\underline{VCR} (Vehicle Traffic Coverage Ratio)}. The ratio of 
traffic  observed by all cameras to the total amount of traffic, which 
%The amount of traffic could be calculated 
%as the  number of position records.
 equals $\sum_{v \in V}T_v(C)/\sum_{v \in V}T_v(R)$.
\item {\underline{VIT} (Vehicle In-Camera Time)}.
The total amount of time
when  a vehicle $v$ is under surveillance, i.e.,  $T_v(C)$
% (i.e., visible to any camera in $C$), which   equals  $T_v(C)$.
\item {\underline{VCH} (Vehicle  Camera-Hits)}.
The number of times that $v$ hits cameras along its trajectory,
%we define vehicle  camera-hits, 
calculated as $\sum_{c \in C}I^\prime_v(\{c\})$, where   
$I^\prime_v(\{c\})$ is
the number of times $v$ hits block 
$c \in C$.
 \item {\underline{VUH} (Vehicle Unique-Camera-Hits)}.
The number of unique cameras  $v$ hits,
 calculated as  $\sum_{c \in C}I_v(\{c\})$, where $I_v(\{c\})$ equals one
 if $v$ hits $c$; otherwise zero.
\end{itemize}

\begin{figure*}[tbh]
	\centering
	\subfloat[S1: top 563  blocks to place cameras  covering all (taxi) vehicles  . \label{fig:beijings1}]{\includegraphics[width=0.48\linewidth]{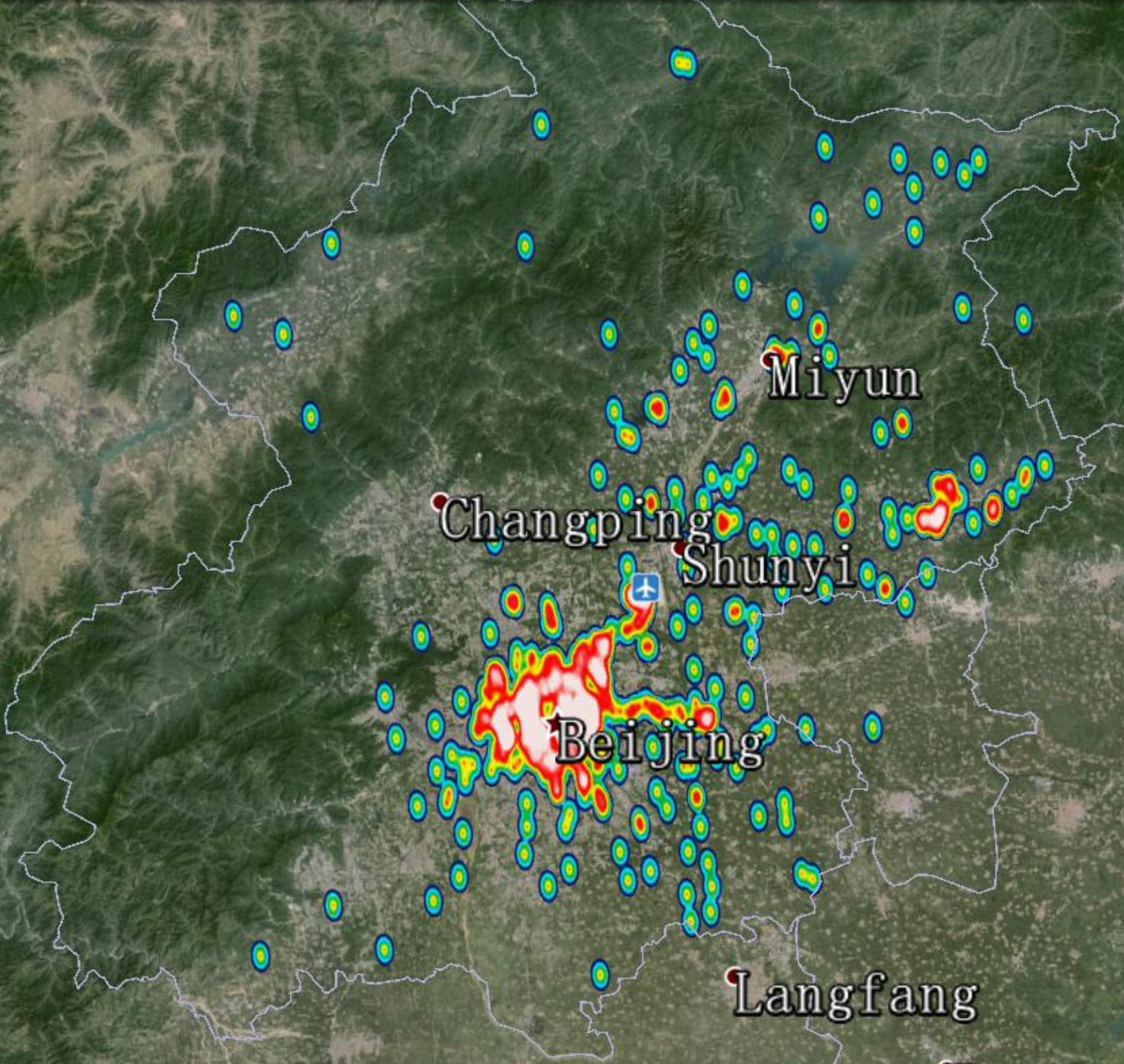}}
	\hfill
	\subfloat[S2: top 563  blocks to place cameras   maximizing VCR. \label{fig:beijings2}]{\includegraphics[width=0.48\linewidth]{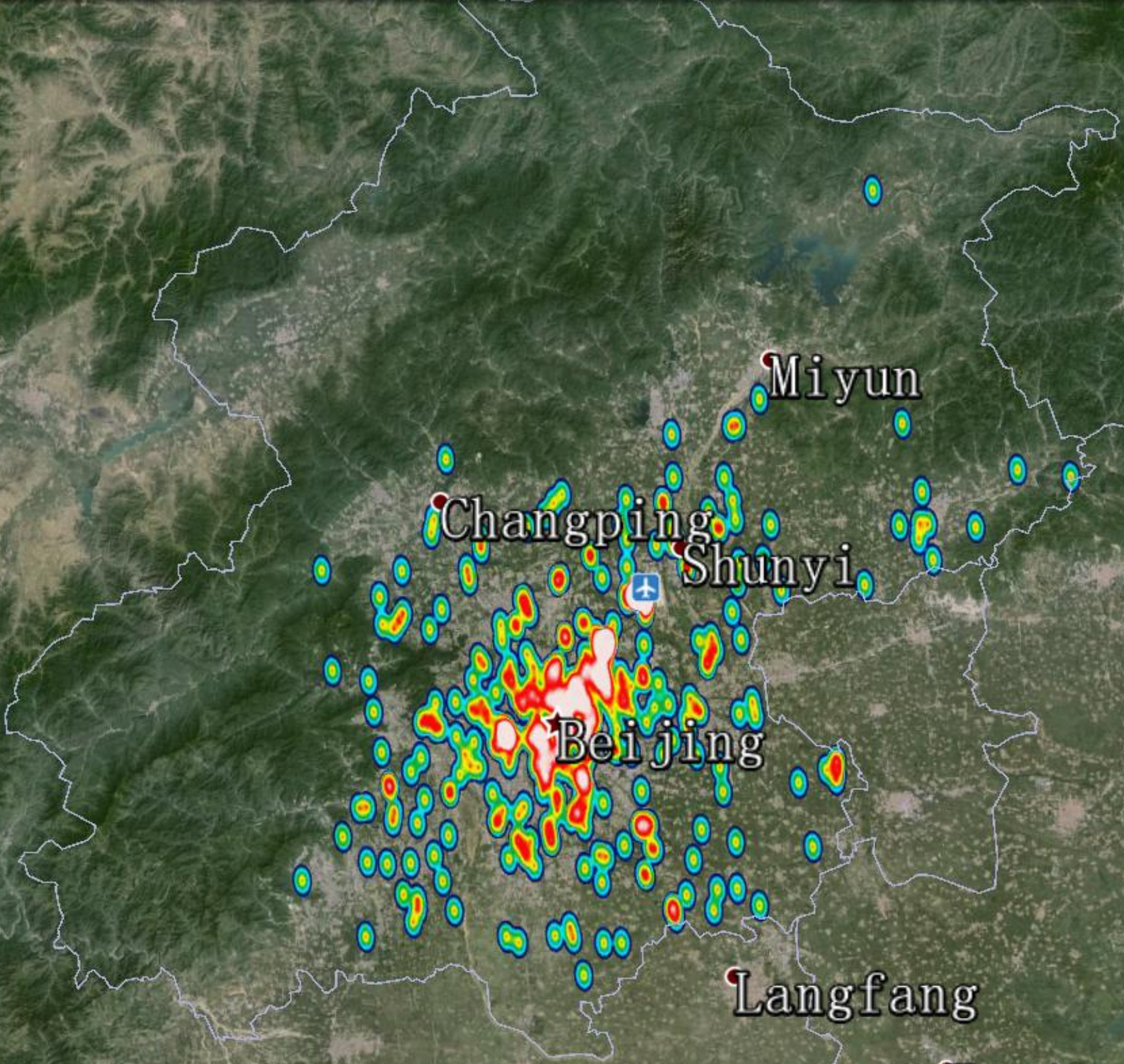}}
	\caption{Example camera placement heatmaps derived using one-week taxi data in Beijing based on strategy S1 and S2.}
	\label{fig2}
\end{figure*}

\subsubsection{\textbf{Result  Analysis}}

\begin{figure*}[tbh]
\centering
\subfloat[UCR   \label{fig:whitelist1}]{\includegraphics[width=0.5\linewidth]{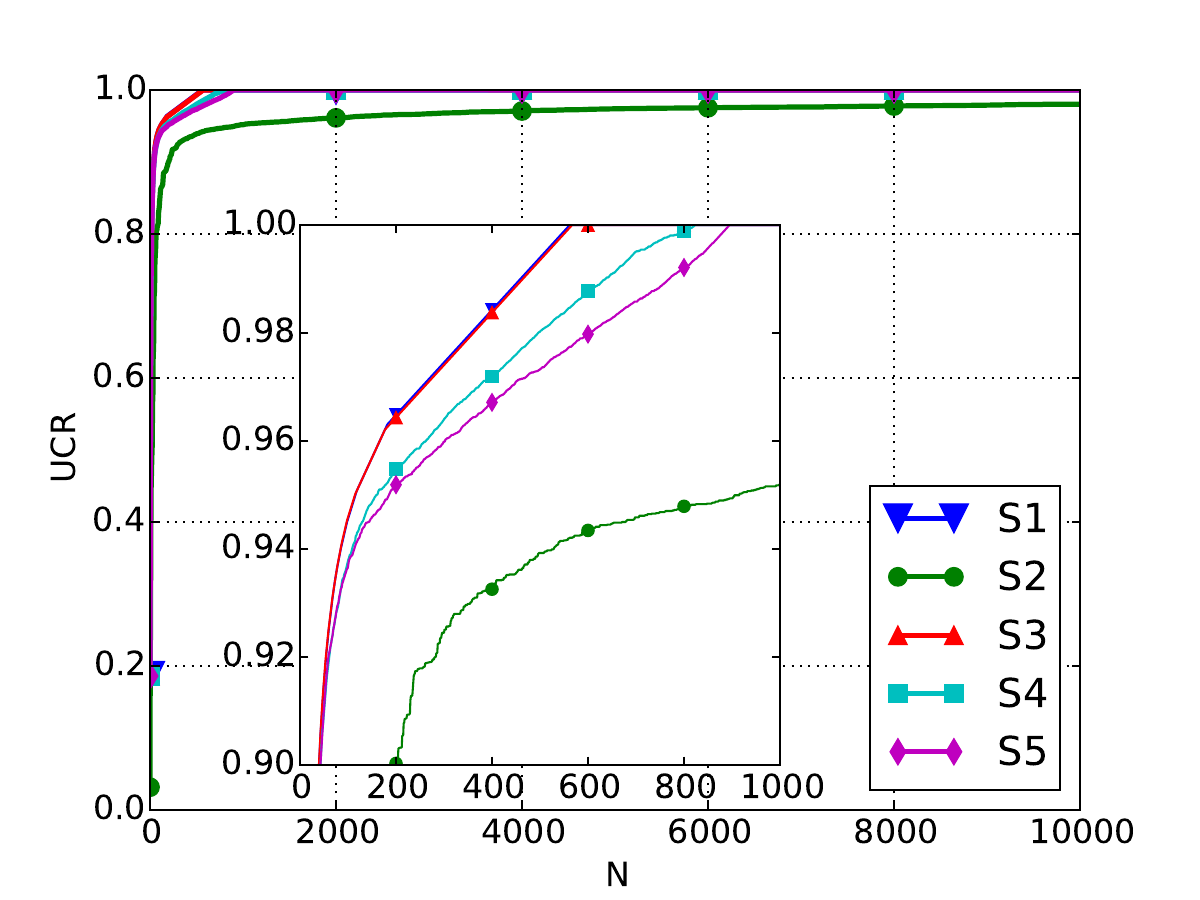}}
\subfloat[VCR  \label{fig:whitelist2}]{\includegraphics[width=0.5\linewidth]{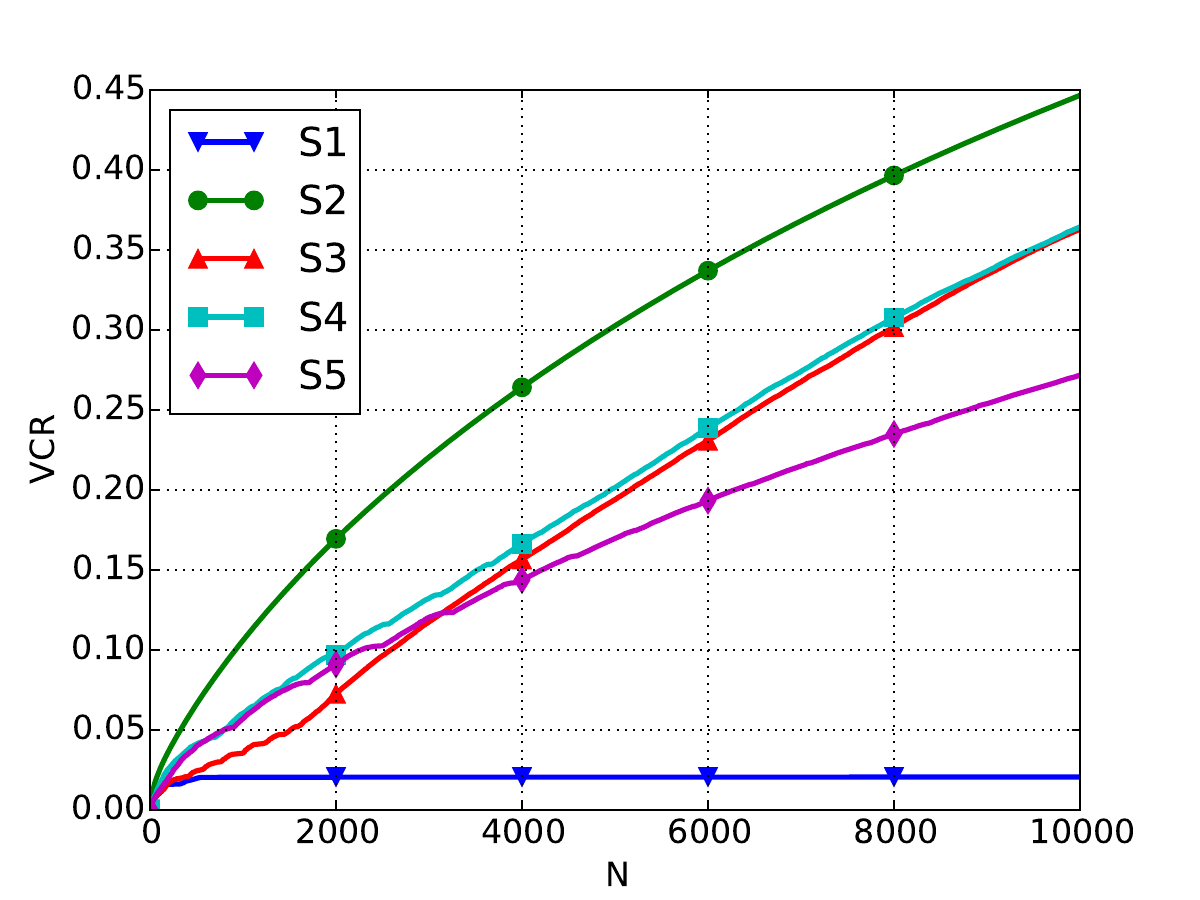}}\\
\subfloat[mean VCH \label{fig:networksize3}]{\includegraphics[width=0.5\linewidth]{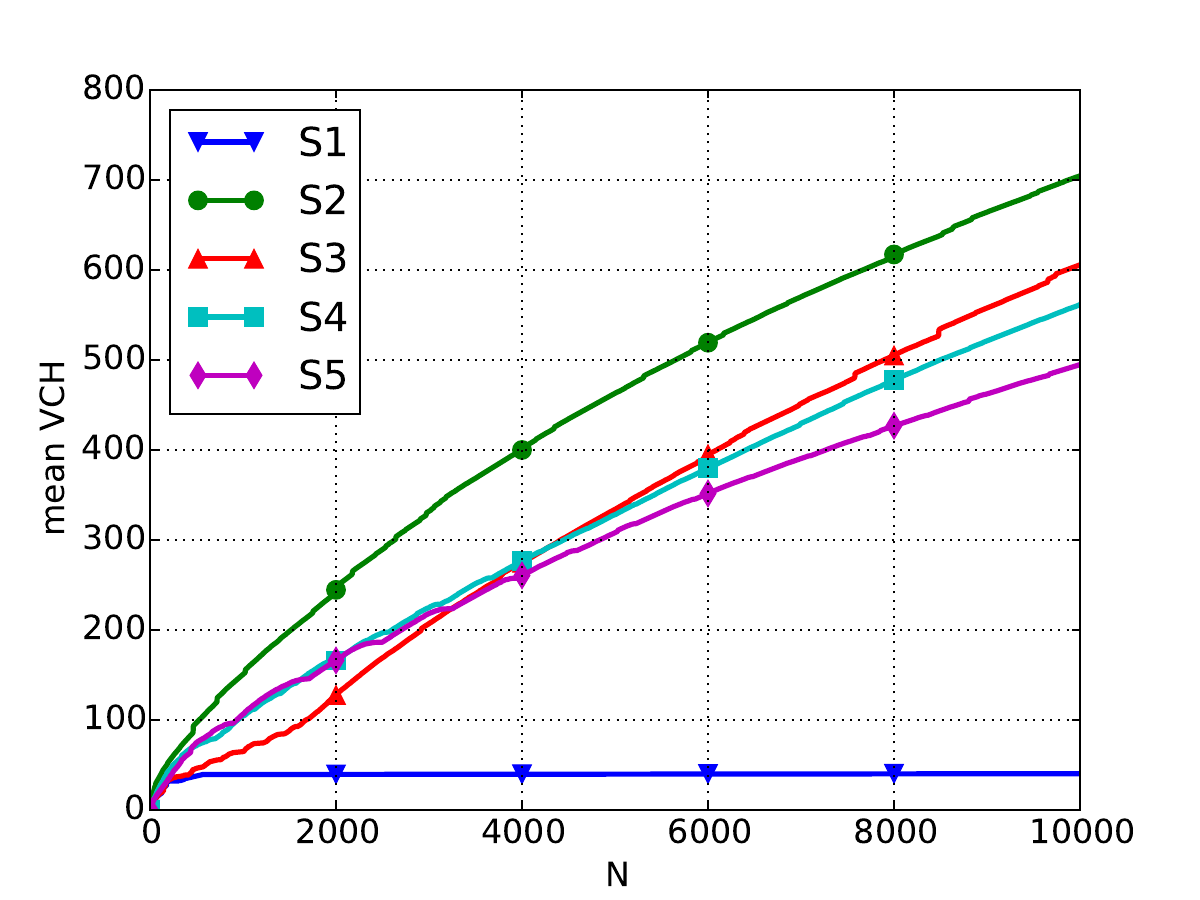}}
\subfloat[mean VUH \label{fig:failure4}]{\includegraphics[width=0.5\linewidth]{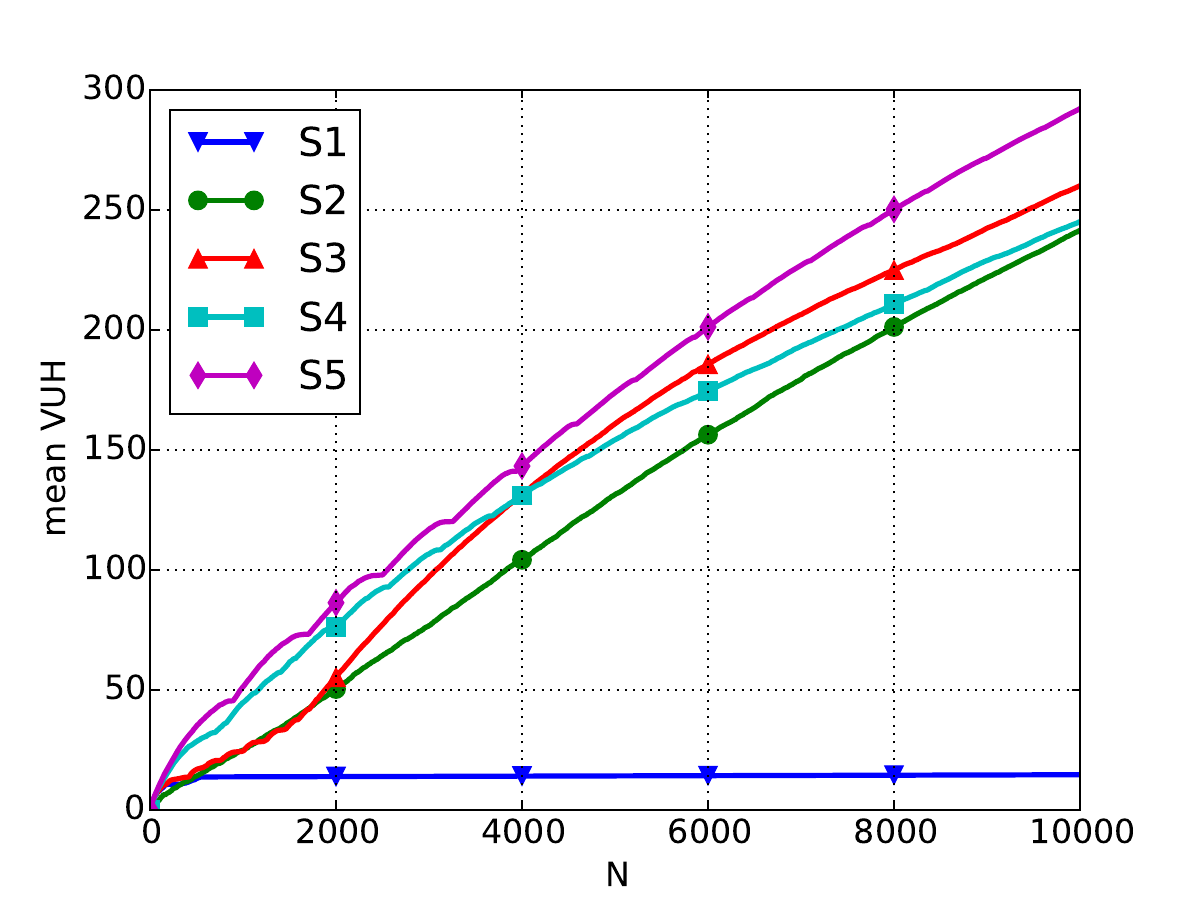}}
\caption{The performance metrics over the number of cameras $N$ of different placement strategies.}
\label{fig1}
\end{figure*}

For each strategy, 
we  calculate UCR,  VCR and VIT by varying the number of  cameras $N$  from 1 to 10,000.
UCR and VCR  reveal
the proportion of unique vehicles and  vehicle traffic   that could be 
covered under each strategy,  respectively.
Additionally, we calculate   VIT, VCH and VUH for each vehicle to have  a closer look at  the total amount time when a vehicle is under surveillance, the total number of  cameras that a vehicle hits, and the total number of unique cameras that a vehicle hits, respectively.

%We increase the  number of placed cameras $N$ from 1 to 10,000.
Fig.~\ref{fig1}  depicts 
the performance   of all strategies.
Generally, the  metrics increase slower   as $N$ becomes larger, meaning that all strategies accomplish  diminishing  returns. 
Particularly, 
all metrics rise  rapidly 
before $N$ reaches 200, accounting for no  more than 0.05\% of    all possible blocks where we can place cameras (i.e., $R$). 
%The resulting performance helps us derive the following observations and insights.

%\subsubsection{Observation 1}

As Fig.~\ref{fig:whitelist1} shows, 
a very small proportion of 
cameras need to be placed to cover the vast majority of vehicles.
Specifically, 
all   strategies exhibit
a rapid rise  in UCR to at least 90\% before $N$ reaches 200, indicating that  we need to place cameras in
no more than 0.05\% blocks to cover at least 90\% vehicles.  
Moreover,
all strategies except S2 achieve a 100\% UCR before $N$ reaches 900 (i.e., 0.2\% of $R$), while S2 almost has no increasing returns after $N=4,000$.
%Therefore, we have the insight below.

\noindent
\textit{\textbf{Insight 1}}.
 \textit{To cover all vehicles,  we need to place cameras 
only in 0.2\% blocks at most using S1, S3, S4, S5.
Among    these strategies capable of
covering all  vehicles,  S1 could cover all vehicles 
with the smallest $N$ (i.e., the most quickly), followed by S3, S4,  and S5 in ascending order.
However,     S2   could hardly converge to cover all   vehicles as   $N$ increases.}

This insight reveals the existence 
of a small set of ``core'' blocks covering all vehicles.
Fig.~\ref{fig:beijings1}
shows the minimum set of  such ``core'' blocks
%to place cameras that cover all  vehicles  during one week
derived from strategy S1.
%, even when $N$ is very large. 
Meanwhile,  $S2$ that only maximizes vehicle traffic coverage  results in slow or even incomplete vehicle coverage.

%\subsubsection{Observation 2}
Fig.~\ref{fig:whitelist2} demonstrates the vehicle traffic coverage ratio, where VCR
rises roughly linearly as $N$ increases for  all    strategies exclusive of S1.
Among all   strategies, 
S2 exhibits the largest growth rate,   followed by S4, S3,  S5, and S1.
Fig.~\ref{fig:beijings2}  shows top 563  blocks where we can place cameras   to  maximize VCR using S2. We observe 
that the blocks in Fig.~\ref{fig:beijings2} are significantly less geographically dispersed
than those in Fig.~\ref{fig:beijings1}.
%S2 is only a little bit larger than S3 that considers the diversity of vehicles, and 
The growth rates of S3 and S4 are comparable.
S1 almost has no increasing returns after $N=200$,
implying that an emphasis on vehicle coverage may fail to maximize traffic coverage.
%as VCR in Fig~\ref{fig:whitelist2}.
%From these observations, 
%we have the insight below.

\noindent
\textit{\textbf{Insight 2}}.
\textit{Although S1 achieves the best performance regarding UCR, its performance regarding VCR degrades drastically to an extent where no  increasing returns is accomplished  when we invest more cameras   after $N=200$. In contrast,  S3$\sim$S5  achieve significant increasing returns regarding VCR as $N$ increases.  Although their increasing returns are not as fast as that of S2,   they can  converge to cover all vehicles quickly while S2 cannot.}

This insight reveals that our proposed  strategies S3$\sim$S5  
can achieve well-balanced performance between UCR and VCR. 
This is because they encourage more  vehicles under surveillance    while simultaneously   considering
more (unique) camera hits and traffic coverage per vehicle.

 Fig.~\ref{fig:networksize3} presents mean VCH across all vehicles, which has similar trend  as VCR. 
 This reflects that  
the number of camera hits is positively correlated to in-camera time. 
For the same reason,  VIT  exhibits   the same trend  as VCH. We therefore omit the plot of mean VIT.   
%The reason is that   the traffic coverage is measured by the  time when a 
%vehicle is in-camera. 
Among all
the strategies, S2
  achieves the best average surveillance performance (i.e., longer in-camera time, more camera-hits) 
per vehicle, though it  may result in slow or even incomplete vehicle coverage. 
Despite the  performance degradation  in mean VCH compared to S2, 
 strategies    
  S3$\sim$S5 achieve   median values of   VIT and VCH across all vehicles comparable to  S2, while simultaneously exhibiting much   smaller
  standard deviation of   VIT and VCH.
%  Fig.~\ref{cdf} depicts
%the distribution of VCH (the same for VIT), where we see less dispersed VCH (and VIT) under  $\text{S3}\sim\text{S5}$. 
%
Furthermore,
% we   calculate  the Gini coefficient of VCH   across all vehicles under strategies $S2 \sim S5$. The Gini coefficient is a measure of statistical dispersion for measuring the inequality among values of a distribution. A Gini coefficient of zero and one means perfect equality and maximal inequality, respectively. The Gini coefficient of VCH for S2 is 0.587, while those for $S3 \sim S5$ equal 0.536, 0.513, and 0.515, respectively. Thus, 
the Gini coefficient of VCH for $S2$  is larger than that for S3$\sim$S5, confirming that  S3$\sim$S5  achieve relatively fairer surveillance across all vehicles regarding in-camera time and camera hits than S2. 
%The insight from Fig.~\ref{fig:networksize3} can be summarized as follows.

\noindent
\textit{\textbf{Insight 3}.
     S3$\sim$S5 achieve  more balanced   surveillance  across all vehicles in terms of  in-camera time and camera hits  than S2, avoiding 
  little surveillance of some vehicles and   too much surveillance (possibly unnecessary) of others.
Therefore, they  provide    relatively fairer security surveillance services across all vehicles.
}    

 Fig.~\ref{fig:failure4} presents 
 the performance of  mean VUH (i.e., the number of unique vehicles that a vehicle hits) across all vehicles.
 We observe that
S5 achieves the largest mean VUH, while S1 has
no increasing returns after $N$ reaches 200.
The remaining strategies exhibit moderate performance worse than that of S5. 
Actually, S5   encourages deploying cameras in the blocks where different blocks along  a vehicle's trajectory can be observed as many as possible, and meanwhile vehicles can be observed as many as possible.
In this case, a vehicle is expected to be observed in more different blocks, thereby beneficial to capturing more places where a vehicle has ever been. 
This is important in security surveillance that prefers the knowledge on the places where a vehicle has been.

\noindent
\textit{\textbf{Insight 4}.
     S5 is a strategy in favor of capturing more places where a vehicle has ever been.
Other strategies perform  worse than S5 regarding mean VUH.  
%     
%      more balanced   surveillance  across all vehicles in terms of  in-camera time and camera hits  than S2, avoiding 
%  little surveillance of some vehicles and   excessive surveillance (possibly unnecessary) of others.
%Therefore, they  provide    fairer security surveillance services across all vehicles.
}    

\subsection{A Case Study of Randomized   Resolution Upgrading}
To demonstrate the performance of randomized resolution upgrading 
based on game theory, 
we consider a security surveillance scenario 
where $N= 10,000$ cameras are placed while only 1,000 cameras could be of high resolution.
% due to bandwidth constraint.  
Suppose the surveillance focuses on criminal (or terrorist) activities 
that tend to be conducted at places that are generally with more vehicle traffic
(so to make more serious consequences).
We thus measure the importance of a block $c_i$ according to its amount of traffic in the dataset, which is calculated as $T(c_i)=\sum_{v\in V} T_v(\{c_i\})$.
%Note that  one can also apply our model in the surveillance that focuses on 
% criminal (or terrorist) activities 
%that tends to be conducted at places with less vehicle traffic
%(so to avoid witnesses).
 
 We   consider that the defender and the adversary play a 
zero-sum game. That is, each player's gain (or loss) of utility is exactly balanced by the losses (or gains) of the utility of the other player(s). 
%In a  two-player zero-sum game,
%the equilibrium of the Stackelberg security game  is equivalent to the  Nash equilibrium
%%, minimax, or maximin strategy
%\cite{Korzhyk:2011:SVN:2051237.2051246}. 
%each player's gain (or loss) of utility is exactly balanced by the losses (or gains) of the utility of the other player(s) (i.e., zero-sum).
 %In other words,  the defender's utility is exactly the  opposite of the adversary's utility.
Formally, we set
$R_i^d+P_i^a=0$, $R_i^a+P_i^d=0$.
Obviously, we have  $U^d+U^a=0$.
Meanwhile, the importance values of  blocks are normalized through dividing each value by the maximum value.
%for simplicity. 
Therefore,  we have 
$R_i^d, R_i^a \in [0,1]$,  and 
  $P_i^d, P_i^a \in [-1,0]$.
  Combining \eqref{uid}, \eqref{uia}, \eqref{object_aduti}, and \eqref{object_deuti},
  it is easy to know $U^a, U^d \in [-1,1]$.

Before realizing randomized resolution upgrading, we randomly generate 1,000 defender pure strategies in total. The union of these pure strategies cover all the placed cameras to ensure that the
resolution of any camera has a chance to be upgraded.  
On the other hand, an adversary  pure strategy   is a block where he commits crimes, the number of which equals 10,000.

Table~\ref{tab:testbed} shows the   defender utility  (i.e., $U_d$) when the defender adopts  the mixed strategy, in comparison to adopting two baseline defender strategies, namely,   uniform strategy and best strategy.  The uniform strategy means the defender adopts a pure strategy uniformly at random.
The best strategy allows the defender to constantly adopt the pure strategy that contains the most important block. If there are multiple such  pure strategies, the defender selects the one maximizing her utility.

%Note that  the  values of the defender utility $U^d$  is negative
%when the defender's security resource investment is limited   (i.e., the number of cameras that can be of high clarity simultaneously). 
%%This is because $U^d$ is an increasing function w.r.t. $x_i$, the marginal probability of protecting a link $i$. 
%Specifically,  in the face of a best response adversary, 
%we have  $U^d=  \sum_{i=1}^{|C|} B_i [x_i R_i^d+(1-x_i)P_i^d]=\sum_{i=1}^{|C|} B_iR_i^d  (2x_i -  1)$. 
%Recall the definition of $B_i$ in \eqref{br}, we derive $U_d=  R^d_{i\_\text{attack}} (2x_{i\_\text{attack}} -  1)$, where ${i\_\text{attack}}$ is the index of the block that the best response adversary selects. 
%Therefore, if $x_{i\_\text{attack}}\geq 1/2$ holds (the marginal probability of monitoring block ${i\_\text{attack}}$ is not less than 1/2), $U^d$ would be non-negative; otherwise negative.  
%As the security resource investment increases, $x_{i\_\text{attack}}$ (i.e., the marginal probability for the defender  to cover $x_i$ across all defender pure strategies)  also tends to increase. 

\begin{table*}[!tbh]
\center
	\caption{The defender's expected  utility against a best-response adversary under different defender strategies (i.e., mixed strategy, uniform strategy, and best strategy) and different camera placement strategies.}
	\begin{tabular}{|l|l|l|l|}		
		\hline		 
		\multirow{3}{*}{\diagbox{\bf placement}{\bf utility}{\bf defender}} & \multirow{3}{*}{\bf mixed strategy} & \multicolumn{2}{c|}{\bf baseline strategies}\\ 
		\cline{3-4}		
		& & 	\multirow{2}{*}{ uniform  strategy} & \multirow{2}{*}{ best strategy} \\ 
		& & & \\ \hline
		S1 &-0.07313   &-0.74600    &-0.19582  \\
		S2   &  -0.13279 &-0.81400 &-0.50654 \\
		S3 &-0.10210   &  -0.77800
   &-0.47290   \\
		S4 &-0.20533   &  -0.81000 &-0.68853   \\
		S5 &-0.12091   &  -0.82800 &-0.49978   \\
		\hline
	\end{tabular}	
	\label{tab:testbed}
\end{table*}

From the table, we see that the mixed strategy achieves significantly larger utility than the baseline strategies, no matter which placement strategy ($S1 \sim S5$) is employed to
 place the    camera infrastructure. 
Such a result indicates that, once the importance of each block is defined, 
one can derive the mixed strategy 
 that outperforms baseline strategies and  maximizes the defender utility against a sophisticated adversary. 
This case study implies the feasibility  that the security police  randomly selects a set of cameras and upgrades their resolution  for high-quality video evidence based on the proposed security game model.

\section{\textbf{Discussion}}

\subsection{\textbf{Customizing Security Surveillance}}
%The types of vehicles may need to be considered in practice.
%The reason is that

Different types of vehicles may 
 require different   surveillance priorities.
 For example,
 compared with individual cars, taxis (as a public space) and school buses usually need higher surveillance priorities. 
 Although 
we do not distinguish between different vehicle 
types, 
every vehicle $v$ can be assigned 
a personalized weight $w_v$ indicating its surveillance priority, 
%One possible solution is to divide the surveillance priorities 
%into five levels, so that  
where larger values of the weight indicate higher surveillance priorities.  
%In this way, a user defined priority could be assigned to individual vehicles.
One can also assign  weights to vehicles
according to their reputation scores derived from social data like   accident records and   vehicle owners'  criminal records.
Vehicles with more accident records, or whose owners have more   criminal records,
would be assigned higher weights.

The weight $w_v$   can be directly used as 
a multiplier of the parameters  in the proposed strategies.   
For example,   $I_v(C)$ could be   replaced by 
$w_vI_v(C)$, and so do 
 $T_v(C)$, $H_v(C)$ and $U_v(C)$.
This
 keeps the non-decreasing submodular property of $F(C)$.
Therefore,   GA remains applicable.
% for
%achieving a solution lower bounded by $63 \%$ of the
%optimal solution.
By incorporating individual  vehicle information, 
we can   customize placement strategies  as needed,  with  a bias toward    vehicles with higher surveillance priorities and lower reputation scores. 

%\subsection{\textbf{Vehicle Reputation}}
%Besides  vehicles types,

\subsection{\bf  \textit{Privacy Issues}}
Determining  strategies for customizing security surveillance 
 not only anticipates GPS data sharing from   vehicles, but also rely on other  social data sources helpful to security surveillance. 
The data sharing may introduce privacy issues, which however
could be resolved by data obfuscation techniques (e.g., obfuscating vehicle IDs). Whenever a security event  occurs, the obfuscated data (e.g., vehicles IDs)  would again be recovered
from  collected  (video)  data  recorded by  cameras  using 
plate number recognition techniques,
after being authorized 
 by privileged  authorities. 
 Additionally, tracking individual vehicles by cameras may also have privacy issues. 
Thus, we suggest the public should be notified of placed cameras, and 
meanwhile the collected video data MUST be strictly managed by authorities according to 
law.

\section{\textbf{Related Work}}

\noindent
\textbf{Transportation Monitoring.}
Transportation monitoring is a broad  area 
closely related to  urban computing tasks like energy and transportation optimization (e.g., ridesharing, speed control),    road traffic monitoring  (e.g., travel time estimation),  urban environment modeling (e.g., PM 2.5 air quality)~\cite{export:211950}. 
Transportation monitoring undoubtedly accelerates 
the operation efficiency of a metropolis.
%Despite 
%vehicles' importance in  carrying 
%massive people and goods   in the operation of a metropolis, 
However,
none of the existing studies regarding
transportation monitoring
exploits vehicle traffic to select camera placement blocks
 in favor of security surveillance.

 \noindent
 \textbf{Physical-World Security Surveillance.}
 Recent years have witnessed a rising trend 
 towards game-theoretic urban security patrolling on roads~\cite{DBLP:conf/aaai/YinAJ14}.
%  ~\cite{tambe_computational_2013,  DBLP:conf/aaai/YinAJ14}. 
%  in the AI community
 Specifically,  an adversary assigns different values to reaching (and
 damaging or destroying) one of multiple targets (e.g., buildings). 
 A defender can allocate resources  to capture
 the  attacker  before  he  reaches  a  target.  
 % In  many  of  these  situa-
% tions, the domain has structure that is naturally modeled as a graph.
% For example, city maps can be modeled with intersections as nodes
% and roads as edges, where nodes are targets for attackers.  
% 
To prevent attacks, security forces can schedule checkpoints on
 roads  to detect adversaries.  
% For instance, in response to
% the devastating terrorist attacks in 2008 [1], Mumbai police deploy
% randomized checkpoints as one countermeasure to prevent future
% attacks. 
Different from these studies, we leverage game theory to control placed cameras. Moreover, camera placement complements road patrolling, because
 the adversaries missed by the road checkpoints 
 could be further inspected by placed cameras.  

\section{\textbf{Conclusion}}
\label{sec:conclusion}

This article explored the linkage between 
vehicle traffic and camera placement in favor of security surveillance from a network perspective. 
We  proposed different   camera placement strategies that facilitate security surveillance.
%To find optimal solutions,
%we developed a greedy algorithm    with a theoretical lower bound. 
Using real-world  data from a metropolis, 
we 
demonstrated that the proposed strategies  could facilitate  metropolis security surveillance  
in different aspects.
We also studied the security surveillance problem that high-resolution  video is desired for high-quality crime evidence, while only a limited number of placed cameras can be  of  high resolution simultaneously based on game theory.
The results illustrated that,
using the carefully designed   randomized strategy,  we  can  
 deter a sophisticated  adversary with maximal utility under the constraint of   practical  overhead.

To make our cities safer, there remain a lot of   problems to explore.
In an era of mass terrorism, we expect our  work could stimulate 
more research on  transportation data-driven security surveillance in smart cities.  
  
%We believe our work 
%%exploring the linkage between 
%%vehicle transportation cyber space  and camera placement strategies in favor of physical-world   security surveillance   from a network perspective. 
%opens up a new opportunity 
%  to reconsider and improve security surveillance in a metropolis. 
%presents
%vehicle traffic driven camera placement,  proposes 
%a diversity of placement strategies, and analyzes how these   strategies could assist in  better  metropolis security surveillance using real-world metropolis traffic. 

%
%% 
% We leave two   research directions as  future work.
% First, we plan to 
%study the performance of our strategies
%from a temporal perspective, and
%develop dynamic strategies difficult  to predict for the adversary.
%For example, when all the cameras cannot be working online concurrently,
%we     develop a well designed randomized strategy that maximizes the surveillance utility, making the adversary feel that the cameras might be working anywhere. 
%  This can  further deter the adversary.
%Second, we will further explore strategies in favor of vehicle trajectory inference from camera data. Inferring the trajectory can help   retrieve  fine-grained  vehicle information critical to security surveillance. 

\balance
\bibliographystyle{unsrt}
\bibliography{sigproc}  % sigproc.bib is the name of the Bibliography in this case

\appendix{{The submodularity for {S3}$\sim$S5 \footnotesize}}

%\section{TEST}

\label{proof-4}
\begin{proof}
\footnotesize

For the  function $F(C)$ defined in {S3},
given every two sets $C_1, C_2 \subseteq \Omega$ and $C_1 \subseteq C_2$, 
we have
\[F(C_1)=\Phi-\sum_{v \in V}  \left(\frac{S}{T_v(C_1)+1}-1\right)/{|V|}\] {and} 
\[F(C_2)=\Phi-\sum_{v \in V}  \left(\frac{S}{T_v(C_2)+1}-1\right)/{|V|}.\] 
{Meanwhile, we add  a new block $c \in \Omega \setminus C_2$ to 
$C_1$ and $C_2$, and obtain}
  \[F(C_1\cup \{c\})=\Phi-\sum_{v \in V}  \left(\frac{S}{T_v(C_1\cup \{c\})+1}-1\right)/{|V|}\] {and} \[F(C_2\cup \{c\})=\Phi-\sum_{v \in V}  \left(\frac{S}{T_v(C_2\cup\{c\})+1}-1\right)/{|V|}.\]
  Accordingly, we obtain the function value increment as
  \begin{align*}
   F(C_1\cup \{c\})&-F(C_1)=\\&\sum_{v \in V}  \left(\frac{S}{T_v(C_1)+1}-\frac{S}{T_v(C_1\cup\{c\})+1}\right)/{|V|}
  \end{align*}
  {and }
  \begin{align*}
  F(C_2\cup \{c\})&-F(C_2)=\\&\sum_{v \in V}  \left(\frac{S}{T_v(C_2)+1}-\frac{S}{T_v(C_2\cup\{c\})+1}\right)/{|V|}.
  \end{align*}
  
Recall $T_v(C)$ denotes 
the amount of traffic contributed by
 $v$ to blocks in $C$. Because  of 
 $c \notin C_1$ and $c \notin C_2$, we have
 $T_v(C_1\cup\{c\})=T_v(C_1)+T_v(\{c\})$, and
  $T_v(C_2\cup\{c\})=T_v(C_2)+T_v(\{c\})$.
  Therefore, 
  the above equations can be rewritten as 
   \begin{align*}
    F(C_1\cup \{c\})&-F(C_1)=\\&\sum_{v \in V}  \left(\frac{S}{T_v(C_1)+1}-\frac{S}{T_v(C_1)+1+T_v(\{c\})}\right)/{|V|}
   \end{align*}
    {and}
    \begin{align*}
     F(C_2\cup \{c\})&-F(C_2)=\\&\sum_{v \in V}  \left(\frac{S}{T_v(C_2)+1}-\frac{S}{T_v(C_2)+1+T_v(\{c\})}\right)/{|V|}.
    \end{align*}

Given $C_1 \subseteq C_2$, 
 we have $T_v(C_1) \leq T_v(C_2) $ holds. Thus, 
      \begin{align*}
    \left(\frac{S}{T_v(C_1)+1}-\frac{S}{T_v(C_1)+1+T_v(\{c\})}\right)\\
    \geq
    \left(\frac{S}{T_v(C_2)+1}-\frac{S}{T_v(C_2)+1+T_v(\{c\})}\right)
      \end{align*}
Finally, we derive 
\[
F(C_1 \cup \{c\})-F(C_1) \geq F(C_2 \cup \{c\})-F(C_2).\]
The submodularity holds for $F(C)$ defined in {S3}.

Similarly, we can deduce that the  submodularity   holds for $F(C)$ defined in S4 and {S5}.
    \end{proof}

\end{document}